# High Field Ultrasound Measurements in UPt$_3$ and the Single Energy Scale Model of Metamagnetism

B.S. Shivaram[1(a)], V.W. Ulrich[2], Pradeep Kumar[3] and V. Celli[1]

[1]Department of Physics, University of Virginia, Charlottesville, VA. 22901

[2] Department of Mechanical Engineering, Grove City College, Grove City, PA. 16127

[3]Department of Physics, University of Florida, Gainesville, Fl. 32611

## ABSTRACT

We report longitudinal ultrasound velocity measurements for magnetic fields up to 33 T applied parallel to the a-axis of the heavy electron compound UPt$_3$. A characteristic dip in the sound velocity at the metamagnetic critical field, $H_c$=20 T, reported in earlier work is reproduced and shown to be independent of temperature at very low temperatures. We show that the single energy scale model (B.S.Shivaram et al., Phys. Rev., **B89**, 241107(R), 2014) captures the observed key features of the field dependence in the sound velocity shift, $δv_s$. The shift $δv_s$ at $H_c$ is found to be inversely dependent on temperature above ~3K and assumes a fixed value at low T. This "saturation" in $δv_s$ below ~3K is accounted for by level broadening of the Uranium spin states.

[(a)]bss2d@virginia.edu – to whom all correspondence should be addressed.

Metamagnetism, the sudden rise in the magnetization at a critical field, is observed in a variety of systems. It can be found in metals[1], correlated oxides[2], polymers[3], and single molecule magnets[4]. From an examination of the universal behavior seen in the magnetic properties of heavy fermion metals we were prompted to propose a single energy scale model[5]. This model captures the behavior of single molecule magnets as well[6]. It is therefore relevant to ask how far such universal aspects apply and inquire whether there are departures (if any) from such universality that define and distinguish the various classes of metamagnets. In the heavy fermion family of metallic metamagnets there is a strong hybridization of the conduction electrons with the local d or f-moments. Thus, in the case of the heavy fermion metal $UPt_3$, for example, while the f-electron in Uranium behaves as a localized electron with a magnetic moment, it assumes at the same time an itinerant character and contributes to observables in the overall electronic response[7]. Similar effects may or may not play out in the other classes of metamagnets. Thus it is useful to explore this dual role of the f-electron or the local spin and establish its universal character in this regard. Experimental probes and concurrent theoretical treatments that assist in this endeavour are particularly valuable.

We explore this theme in this paper through high resolution sound velocity measurements in high magnetic fields on a single crystal of $UPt_3$. The sound velocity has a minimum at the metamagnetic critical field that depends inversely on temperature for T > 3K. While there have been previous reports on ultrasound measurements in $UPt_3$ for magnetic field parallel to the basal plane we identify a saturation over the entire field range in the sound velocity change in the very low temperature limit.

These experimental results can be understood within the framework of a single energy scale model for metamagnetism[5,8]. In a model metamagnet the quantum spin is in a singlet ground state separated from a doublet excited state. The latter is split by the magnetic field and the lower energy level of the doublet crosses the non magnetic ground state at a critical field. In analogy with the scaling ansatz of Fulde and Thalmeier[9,10], we allow for a spin lattice interaction and the strain dependence of the energy level separation.  As we see below this elucidates many of the experimental results.  By introducing a (temperature independent ) broadening of the energy levels, which scales with the single energy further improvement of the model to match experimental observations is obtained.

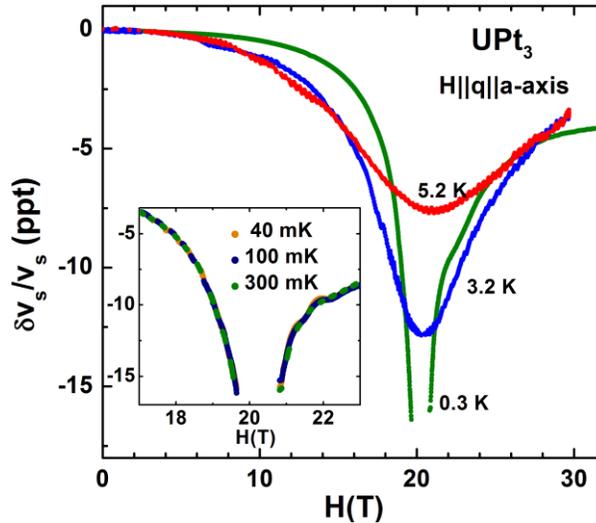

*Figure 1: Shows the change (in parts per thousand) in the longitudinal sound velocity for magnetic field along the a-axis of $UPt_3$ for five different temperatures. While the results shown in the main part of the figure indicate a significant temperature dependence to the width of the dip, the inset demonstrates that this dependence vanishes at the lowest temperatures.*

The measurements we report were performed at the National High Magnetic Field Laboratory, Tallahassee, in a 33 T Bitter magnet. We used the same high quality single crystal of $UPt_3$ that was employed in earlier measurements of the superconducting phase diagram under uniaxial stress[11]. In fig.1 we show the changes in the longitudinal sound velocity that occur through the metamagnetic transition in $UPt_3$ centered at 20 T.  Such large changes in the sound velocity are normally observed in the vicinity of structural phase transitions.  Instances where an electronic effect gives rise to changes in the elastic constants of this magnitude are rare.  The results shown in fig.1 are consistent with the previous work carried out by the Frankfurt group[12] (q||b-axis) and the Milwaukee-Northwestern collaboration for q||H||b-axis[13] except that in our case both the magnetic field and the sound wave vector were parallel to the a-axis.  The salient features in this figure are the gradual sharpening of the response as the temperature is lowered and the concomitant rapid increase in the velocity dip at the critical field as shown in the main part of the figure.  In contrast the inset shows the observed very low temperature response in the vicinity of the transition with no discernible temperature dependence. There is a weak dip or 'knee' on the high field side at 22 T.  We ignore this particular feature in the present paper.

We next examine the data of fig.1 in the context of the single energy scale model.  In this model metamagnetism is described by an effective Hamiltonian, with the three eigenvalues 0, (Δ+γH) and (Δ-γH).  Here Δ sets the energy scale and is the separation between the singlet ground state and the excited doublet, H is the magnetic field and γ is the magnetogyric ratio.  If the sample consists of N independent units with this level scheme, the free energy can be written as[14] $F_0(\tau) + Nf(\tau, b)$, with

$$f(\tau, b) = -\tau\Delta \ln\left[1 + 2e^{-1/\tau}\cosh\left(\frac{b}{\tau}\right)\right] \quad (1)$$

Here and in the following we use the dimensionless variables $b = \frac{\gamma H}{\Delta}$, $\tau = \frac{k_B T}{\Delta}$ and $F_0(\tau)$ is the free energy of the background lattice.  In the range of frequencies we work with (20 MHz to 60 MHz) the propagation of sound is nearly isothermal.  The field-induced change in the velocity of sound $\delta v_s$ is related to the elastic constant change δc by $\frac{\delta v_s}{v_s} = \frac{1}{2}\frac{\delta c}{c}$.  The elastic constant change is defined as the second derivative of the free energy per unit volume with respect to strain ε.  Then

$$\delta c = \frac{N}{V}\left[\frac{\partial^2 f}{\partial \Delta^2}\left(\frac{\partial \Delta}{\partial \varepsilon}\right)^2 + \frac{\partial f}{\partial \Delta}\frac{\partial^2 \Delta}{\partial \varepsilon^2}\right] \quad (2)$$

The quantity $\left(\frac{\partial \Delta}{\partial \varepsilon}\right)$, a constant, is experimentally accessible through the uniaxial strain dependence of the metamagnetic critical field, $H_c = \frac{\Delta}{\gamma}$. From Bakker, de Visser, Menovsky, and Franse[15] we find $\left(\frac{\partial \ln H_c}{\partial \ln V}\right) = 59$, and we can take this to be the same as $\left(\frac{\partial \ln \Delta}{\partial \ln V}\right)$ since we have already assumed in (2) that γ is strain independent.  Further, in the model Δ =1.5 $k_B T_1$, where $T_1$ =20 K is the temperature where a peak in the linear susceptibility occurs[5].  From measurements of $T_1$ under pressure by Willis et al. [16], $\left(\frac{\partial lnT_1}{\partial P}\right) = 25$ Mbar$^{-1}$. Using the measured compressibility[17] of UPt$_3$ of 0.48 Mbar$^{-1}$ the latter figure yields $\left(\frac{\partial ln\Delta}{\partial lnV}\right) = 52$. Thus both methods yield comparable values[18]. If we neglect (at first) the term $\frac{\partial^2 \Delta}{\partial \varepsilon^2}$ in (2) the fractional change in the elastic constant becomes:

$$\delta c/c = -(N\Delta/Vc)\left(\frac{\partial ln\Delta}{\partial \varepsilon}\right)^2 \frac{2e^{1/\tau}\cosh(b/\tau)}{\tau(2\cosh(b/\tau)+e^{1/\tau})^2} \quad (3)$$

From this expression, for instance the maximum shift in $v_s$ (i.e. the minimum in $\delta v_s/v_s$) at 3.3 K is predicted to be -50 ppt of the same order of magnitude as the value shown in fig.1. Confining our attention to the vicinity of the critical field, b=1, it is easy to see from (3) that in the T→0 limit $\delta v_s$ varies as 1/4T. Figure 2 shows the experimental values of $\left[\frac{dv_s}{v_s}\right]^{-1}$ at the critical field plotted against temperature. The linear temperature dependence predicted by the model is apparent. However, a more careful examination of the very low temperature behavior is necessary.

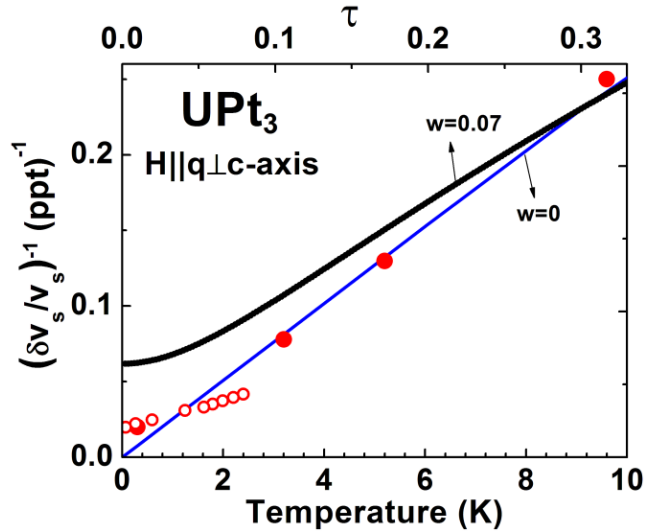

*Fig.2: Shows the inverse of the measured total change in the sound velocity (solid dots) at $H_c$ in UPt$_3$. The blue line is the $T^{-1}$ behavior from the model eq.(3). The solid dark line is the calculated response with the energy levels broadened (see text). The open red circles are from ref. 9.*

In fig. 3 we show the full field dependence of the elastic constant changes obtained from (3). The inset in this figure shows the predicted narrowing of the width at the lowest temperatures, a feature clearly absent in the experimental data (inset of fig.1). Instead, the observed response is "saturated". In addition the experimental data is not symmetric about the critical field $H_c$. We note that similar saturation and asymmetry effects may be seen in earlier data for UPt$_3$ (refs.12,13) as well as for CeRu$_2$Si$_2$ [19]. These features are also readily explained within the context of our model.

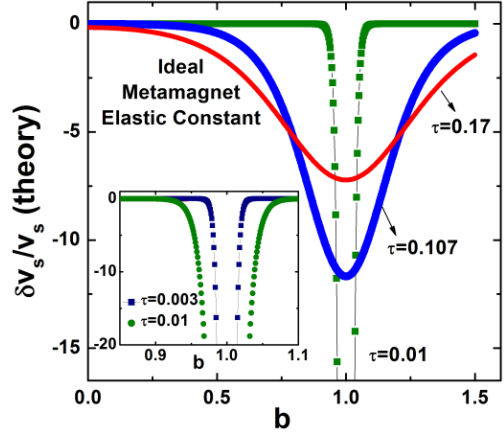

*Fig.3: Shows the elastic constant changes as derived from the thermodynamic relation (4) in the model. The inset shows two very low temperature responses (corresponding to 300 mK and 100 mK for UPt$_3$). A narrowing is predicted even for such low temperatures for an ideal metamagnet.*

In order to do so we introduce a broadening of the energy levels of the effective spin and also take into account the second term in (2) proportional to $\frac{\partial^2 \Delta}{\partial \varepsilon^2}$. The physical origin of the broadening can be due to a number of reasons, including the coupling of the local moment to the conduction electrons[20]. The effect of such a coupling essentially is to replace the temperature $\tau$ by $\sqrt{(\tau^2 + w^2)}$ in the expression for f (eq.1), where 'w' plays a role akin to the

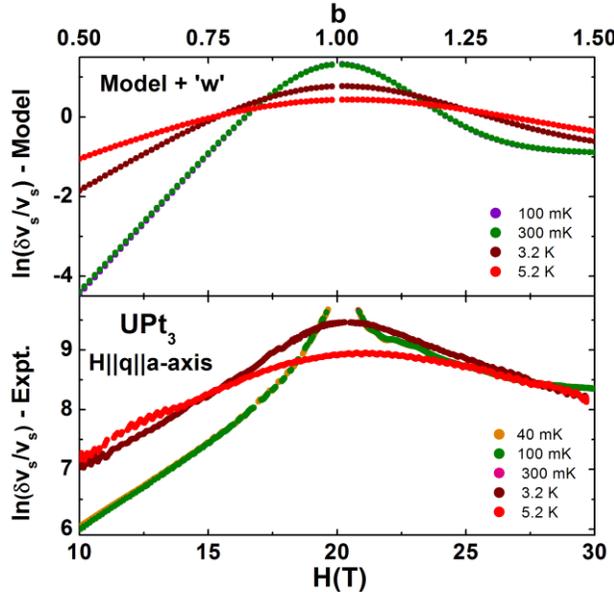

*Figure 4: Shows the very low temperature experimental sound velocity plotted on a log scale as a function of the magnetic field, lower panel. The top panel presents model calculations with the level broadening parameter w=0.07. Contrary to the inset of fig.3 the calculated response is identical at the two lowest temperatures shown.*

Dingle temperature. With this modification we have evaluated δv$_s$/v$_s$=1/2 δc/c using eq.(2) with $\Delta \frac{\partial^2 \Delta / \partial \epsilon^2}{(\partial \Delta / \partial \epsilon)^2}$ taken as an adjustable parameter "a" (asymmetry parameter). The evaluated results with a=-0.4 are presented in fig.4.  The deep variation of the sound velocity at the lowest temperatures found in the inset of fig. 3 is absent and the response at very low temperatures is essentially identical. To generate the results in fig.4 we used the factor w=0.07 to obtain good fits with the experimental results. This value of w corresponds in temperature to a value of 2 K. This is also the temperature where a deviation from the 1/T behavior of the maximum sound velocity dip sets in as evident from fig.2 and consistent with the data presented by Feller et. al[9]. We also show in fig.2 (black line) the predicted temperature dependence of the sound velocity dip at b=1.  It is clear from this line as well as by examining the full field dependence in fig. 4 that the model underestimates the sound velocity change in the vicinity of the critical field: there is a "sharpening" of the ultrasound velocity immediately near $H_c$.  This discrepancy can perhaps be understood in terms of the need to treat separately the different components of the stress tensor, and not to assume simply that they all scale with pressure. The data are taken at constant pressure and no other stress, and that is not the same as constant volume and no strain, the conditions that apply to the model. Perhaps an analysis of the effects of magnetostriction can be performed along the lines presented by Matsuhira et al[21], but that would involve additional assumptions and parameters.  The "sharpening" could also be due to the enhancement of ferromagnetic correlations that emerge at the metamagnetic transition as noted in several previous studies[22].

In conclusion, we have presented results from longitudinal sound velocity measurements in $UPt_3$ for magnetic field, H||a-axis.  We have also shown that the recently proposed single energy scale model of metamagnetism augmented to broaden the local moment energy levels is able to provide a good quantitative description of the observed features.  Through this model it is also possible to obtain the temperature dependence of the sound velocity in zero field.  The model results are similar to the measurements by Yoshizawa et. al.[23] over a broad temperature range and also reproduce the significant temperature dependence observed below 1 K[24,25,26]. This sub-kelvin temperature dependence of the sound velocity together with similar behavior of the thermal expansion coefficient is generally attributed to the Fermi liquid nature of $UPt_3$.  Thus the dimensionless parameter "w" (when assumed to be given by W/Δ with W the level broadening) could indeed be a measure of the local moment – conduction electron interaction i.e. the width of the Kondo resonance[20]. With W having the same pressure dependence as Δ it is easily seen that the empirical scaling[27] $\chi(0)^{-1} \sim \Delta$ is also preserved.  Since sound velocity changes compared to other physical quantities can be tracked very precisely they present an outstanding opportunity to explore itinerant metamagnetism and compare the results with model calculations as demonstrated here.

**Acknowledgements:**  It is our great pleasure to thank Eric Palm, Tim Murphy and Scott Hannah of the NHMFL, Tallahassee, Florida, for their kind help during these measurements. The work at the University of Virginia was supported through NSF DMR 0073456. We thank David Hinks for the single crystal of $UPt_3$ and acknowledge useful conversations with Tony Leggett.